# Drops deformation and magnetic permeability of a ferrofluid emulsion

Arthur Zakinyan, Yury Dikansky

Department of General Physics, Stavropol State University, 1 Pushkin St, 355009 Stavropol, Russia

Address correspondence to A. Zakinyan:
phone: +7-8652-57-00-33,
postal address is presented above,
e-mail: zakinyan.a.r@mail.ru

**Abstract.** In the paper the novel soft magnetic composite system is investigated. A ferrofluid emulsion studied demonstrates the strong magnetic properties which are atypical for commonly known emulsions. Interaction of ferrofluid emulsions with a magnetic field is considered. Structural transformations in these media, such as deformation of emulsion microdroplets and emulsion inversion, are studied. The changes in the relative permeability of emulsion associated with structural transformations are investigated. The theory of the observed phenomena is developed, and the feasibility of effectively controlling the magnetic properties of ferrofluid emulsions by applying a magnetic field is demonstrated.

## 1. Introduction

The use of the materials which can become anisotropic under the external action is widely considered as an effective way to obtain desirable material properties. One of the approaches to achieve this goal is based on the use of composite material, whose properties strongly depend on its structural microgeometry. In several researches much attention has been paid to the use of external electric fields for the modification of properties of a composite material [1–3]. The magnetic-field-induced structural transformations are also observed in different soft magnetic systems [4, 5]. As an example, in several recent works, attempts have been undertaken to create new liquid magnetizable composites based on ferrofluids that can more effectively interact with external magnetic fields [6–8]. In the present work we study magnetic and structural properties of new composite system which is a ferrofluid emulsion.

A ferrofluid is a colloidal suspension of ultra-fine ferro- or ferri-magnetic nanoparticles suspended in a carrier fluid. Previously some properties of the emulsions with ferrofluid droplets dispersed in water have been studied in several works. The droplets of the studied emulsions hold the spherical shape and align in chain-like aggregates under the action of magnetic field. The structural [9, 10], optical [11], and magnetic [12] properties of such emulsions have been investigated. In this paper we consider the ferrofluid emulsions with deformable droplets under the action of magnetic field. Furthermore, we will consider the emulsions comprising a ferrofluid as a dispersed phase and also as a dispersion medium.

As reported in [13, 14], ferrofluid with drop-like aggregates exhibits similar properties. Such system was obtained by diluting the initial kerosene-based ferrofluid with a coagulator mineral oil. Drop-like aggregates are easily deformable in magnetic fields, but the concentration of aggregates in ferrofluid cannot be controlled. In addition, it is difficult to determine the properties of aggregates and surrounding medium independently and precisely. The ferrofluid emulsions studied in the present work do not possess such disadvantages.

## 2. Experimental

In our experiments we used a kerosene-based ferrofluid with dispersed magnetite nanoparticles of about 10 nm diameter stabilized with oleic acid. The properties of the ferrofluid are: density is 1620 kg m$^{-3}$, dynamic viscosity is 29 mPa s, magnetite volume fraction is 19% and saturation magnetization is 55.4 kA/m. The ferrofluid used was made by OJSC "NIPIgaspererabotka" (Russia).

Ferrofluid emulsion may be produced by dispersing either a ferrofluid in a nonmagnetic liquid or a nonmagnetic liquid in ferrofluid. FH 51 aviation oil immiscible with the ferrofluid was selected as the nonmagnetic liquid. Its density is 776 kg m$^{-3}$ and the dynamic viscosity is 14.5 mPa s. The main reason to use this oil is that the interfacial tension at the interface between it and the ferrofluid is quite low ($\sigma = 10^{-6}$ N/m). The radii of the emulsion droplets are varying from 1 to 5 μm, so the ferrofluid can be considered as continuous liquid magnetizable medium. No stabilizing agents were used in the preparation of the emulsions.

The structural state of the emulsion was examined under an optical microscope. It was found that, when emulsion is subjected to a magnetic field, its drops can deform and take the shape of ellipsoid of revolution elongated along the direction of a magnetic field. Because of low interfacial tension, under the action of comparatively weak magnetic field significant deformation of droplets takes place. As an example, the layer of emulsion of oil drops in ferrofluid under the action of constant uniform magnetic field produced by the Helmholtz coils ($H = 1.5$ kA/m) is shown in Fig. 1. The volume fraction of oil drops is 10%. We have measured the semiaxial ratio $a/b$ ($a$ is the major semiaxis parallel to the field and $b$ is the minor semiaxis) of the deformed oil drops in this emulsion as a function of the magnetic Bond number defined as Bm=$\mu_0 H^2 R/\sigma$, where $\mu_0$ is the magnetic constant, $H$ is the magnetic field strength, and $R$ is the radius of the drop in zero field. The results are shown in Fig. 2.

It was observed that, if an emulsion is concentrated to ~ 50 vol. %, it will invert when diluted with additional internal phase. These structural transformations must change the macroscopic properties of the emulsion, in particular its magnetic properties.

The relative permeability of the emulsion was determined by measuring the inductance of a solenoid with the emulsion inside. In our experiments we used the measuring bridge "LCR meter-817" to measure the inductance. The frequency of a measuring field was 1 kHz and its strength was sufficiently small (~ 50 A/m) to have no effect on the structural state of the emulsion. To study the influence of a magnetic field on the permeability of the ferrofluid emulsion, it was subjected to an external constant magnetic field directed in parallel with the alternating measuring field. To this end, the solenoid filled with the emulsion was placed in a cubic magnetizing system providing a uniform magnetic field at the place of the solenoid. Since the diameter of the solenoid (0.5 cm) is much less than its length (20 cm) we can assume that the magnetic field inside the solenoid is equal to the external magnetic field.

Fig. 3 shows the obtained dependences of the emulsion permeability on the constant magnetic field strength at different volume fractions of the dispersed phase. Squares in Fig. 3 correspond to the emulsion of oil droplets dispersed in the ferrofluid (hereafter, oil-in-ferrofluid emulsion); circles correspond to the emulsion of ferrofluid droplets dispersed in the oil (ferrofluid-in-oil emulsion). As is seen, these dependences have maxima, which are not observed for the initial pure ferrofluid and may serve as a characteristic peculiarity of ferrofluid emulsions. Similar field dependence of magnetic permeability of ferrofluid with drop-like aggregates was observed in [13, 14]. The field dependence of the permeability of the pure ferrofluid, $\mu_f$, used in our experiments is shown in Fig. 4.

The appearance of a maximum can be explained in the following way. The influence of a magnetic field on the permeability of a ferrofluid emulsion shows up in the form of two competing factors. Upon stretching the microdrops in a constant field the demagnetizing factor diminishes for each drop and the magnetic permeability of an emulsion grows. On the other hand, the pure ferrofluid decreases in permeability when the biasing field is augmented, as follows from Fig. 4.

As is seen from Fig. 3, the run of the curves is nearly the same; however, the permeability of ferrofluid-in-oil emulsion rises more considerably in the low-field range. It may be assumed that the effect of drop elongation and the decrease in the demagnetizing factor are more pronounced for ferrofluid drops dispersed in the oil. It should be noted that weaker deformation of nonmagnetic drop in magnetic medium in comparison with the magnetic drop for the same values of the parameters was previously observed in [14].

The concentration dependence of permeability of an emulsion has been measured. The amount of the dispersed phase was increased by gradually raising the volume fraction of the ferrofluid in the emulsion. The obtained dependence of emulsion permeability on the volume fraction of the ferrofluid is presented in Fig. 5. In the absence of an additional magnetic field, the concentration dependence is seen to have an inflection point at concentration ~ 50 vol. %, which indicates a change in the phase state of the emulsion. At this concentration phase inversion from ferrofluid-in-oil emulsion to oil-in-ferrofluid takes place. The action of magnetic field changes the run of the concentration dependence in Fig. 5. In particular, the inflection point virtually disappears. Thus, it may be concluded that inversion of the emulsion shows up as a significant change in the character of its magnetization.

Each dependence of magnetic permeability has been measured three times, and the obtained results are within 5% of each other.

It should be noted that the sedimentation of droplets can take place in the studied emulsions. Microscopic observations and magnetic measurements have shown that marked changes of properties of the emulsion owing to sedimentation can be detected in 1 or 2 h after emulsion preparation. On the other hand, it takes several minutes to perform each measurement described above. Thus we can consider an emulsion stable during measurements and neglect the sedimentation in the further analysis.

## 3. Theory

Theoretical analysis of magnetic permeability of a ferrofluid with drop-like aggregates in effective medium approach is given in [14]. Wagner's formula valid in a limit of low concentrations was used for calculation of permeability. In the present work we extend this analysis to the case of high volume fractions by using Bruggeman's formula for the effective medium approximation. Also we take into account the effect of macroscopic magnetic anisotropy of the emulsion.

In consequence of droplets deformation under the action of magnetic field the emulsion becomes anisotropic. We will assume that emulsion droplets are identical and have a shape of a prolate ellipsoid of revolution with the semiaxial ratio $\gamma = a/b$. As shown in [15], the magnetostatic problem for an ellipsoid characterized by a $\gamma$, embedded in an anisotropic medium, can be transformed into the problem for an ellipsoid with a renormalized $\gamma' = \gamma(\mu_x/\mu_z)^{1/2}$ in an isotropic medium. Taking into account the macroscopic anisotropy of the medium, the components of $\mu$ are determined by the system of nonlinear equations:

$$\frac{\mu_z - \mu_i}{\mu_e - \mu_i}\left(\frac{\mu_e}{\mu_z}\right)^{N_z} = 1 - \varphi \qquad (1)$$

$$\frac{\mu_x - \mu_i}{\mu_e - \mu_i}\left(\frac{\mu_e}{\mu_x}\right)^{N_x} = 1 - \varphi \qquad (2)$$

where $\mu_i$ is the permeability of internal phase, $\mu_e$ is the permeability of external phase, $\varphi$ is the volume fraction of internal phase, and

$$N_z = \frac{1-e(\gamma')^2}{2e(\gamma')^3}\left(\ln\frac{1+e(\gamma')}{1-e(\gamma')} - 2e(\gamma')\right) \qquad (3)$$

$$N_x = \frac{1-N_z}{2} \qquad (4)$$

are the demagnetizing factors of particles of internal phase [16], with $e(\gamma)=(1-1/\gamma^2)^{1/2}$ being the eccentricity. The direction of a magnetic field corresponds to the $z$ axis. The renormalized semiaxis ratio $\gamma' < \gamma$ takes into account the effect of macroscopic magnetic anisotropy.

The value of $\mu_z$ has been measured in the experiments described above. According to (1) and (2), the permeability of the emulsion nonlinearly depends on the volume fraction of the dispersed phase. The permeability also depends on the external magnetic field, since demagnetizing factor, $N_z$, decreases with increasing field due to deformation of the droplets. To find the demagnetizing factor as a function of the field strength, we need to determine the relationship between the droplet deformation, $\gamma$, and the magnetic field strength. The problem of deformation of single ferrofluid drop under the action of magnetic field has been investigated experimentally and theoretically in several works; the review of these works is presented in [14]. The results of [14] are valid only in the limit of a small volume fraction, $\varphi$, of embedded droplets, when the magnetic interaction between them can be neglected. In this work we extend the results of [14] to the case of finite-$\varphi$ composite materials by considering the magnetic interparticle interaction, which becomes significant on increase of $\varphi$. We will consider the competition between the magnetic energy and the interfacial energy. The equilibrium shape of each droplet corresponds to the minimum of the free energy of a unit volume of an emulsion. Within the effective medium approach the free energy, $F$, of a unit volume of monodisperse emulsion can be written as

$$F = -\frac{1}{2}\mu_0(\mu_z - 1)H^2 + \frac{\varphi}{V_1}W_{s1} \qquad (5)$$

where $V_1 = 4/3\pi R^3$ is the volume of emulsion droplet, $\mu_0 = 4\pi\,10^{-7}$ H/m, $W_{s1}$ is the surface energy of individual drop. We will neglect the magnetostriction effect and assume that $V_1$ is constant. Surface energy of an ellipsoidal drop is given by

$$W_{s1} = 2\pi\sigma R^2\left(1-e(\gamma)^2\right)^{1/3}\left[1 + \frac{\arcsin e(\gamma)}{e(\gamma)\left(1-e(\gamma)^2\right)^{1/2}}\right] \qquad (6)$$

where $\sigma$ is the interfacial tension and $e(\gamma)=(1-1/\gamma^2)^{1/2}$. The equation determining the position of the free energy, $F$, extremum can be written as

$$\partial F/\partial \gamma = 0 \qquad (7)$$

The relationship between the drop deformation and the field strength may be obtained from numerical analysis of (7). Fig. 6 shows the calculated dependence of the droplet deformation $\gamma$ in a ferrofluid-in-oil emulsion on the dimensionless parameter Bm for different values of $\varphi$. The magnetic Bond number, Bm, characterizes the relative strength of the magnetic and surface forces. The values $\mu_i = 6.4$ and $\mu_e = 1$ were used for calculations, presented in Fig. 6. As is seen, the deformation depends strongly on the volume fraction. A comparison of the calculated deformation of oil-in-ferrofluid emulsion droplets with the experimentally observed deformation is presented in Fig. 2.

Thus, the solution to the problem of magnetic permeability of a ferrofluid emulsion is presented by expressions (1) – (7). It is necessary to take into account that the permeability of one of the emulsion components (ferrofluid) is also field-dependent. We will employ the empiric field dependence of the ferrofluid permeability, shown in Fig. 4. The problem can be solved numerically. When describing the behavior of the oil-in-ferrofluid emulsion, $\mu_e$ denotes the permeability of the ferrofluid, $\mu_f$, and $\mu_i = 1$. For the ferrofluid-in-oil emulsion, on the contrary, $\mu_e = 1$ and $\mu_i$ is the permeability of the ferrofluid. Drop mean radius $R \approx 3$ μm was used for calculations. The results of calculations are shown in Figs. 3 and 5. As is seen, experimental results are in good agreement with theoretical calculations.

The essentially different responses of the ferrofluid-in-oil and oil-in-ferrofluid emulsions to application of magnetic field are observed in the concentration dependence of the relative change of the emulsion permeability (Fig. 7). In Fig. 7 curve *1* is the theoretical dependence calculated for the oil-in-ferrofluid emulsion; curve *2* is the theoretical dependence for the ferrofluid-in-oil emulsion. Both dependences are plotted for the case when the permeability changes under the action of field $H = 750$ A/m. This value of a magnetic field approximately corresponds to the maximum on the field dependence of emulsion permeability. For comparison, the data points are also shown. The ferrofluid-in-oil emulsion is seen to respond to the field more strongly. The smaller effect in the case of oil-in-ferrofluid emulsion is the direct consequence of weaker deformation of nonmagnetic drop in ferrofluid in comparison with the ferrofluid drop in nonmagnetic liquid.

Fig. 7 also demonstrates the change of magnetic properties of an emulsion during the phase inversion process. From Figs. 5 and 7, it follows that in the concentration region from 40 to 60 vol. % (approximately), the emulsion exhibits intermediate properties. In this concentration range the emulsion properties gradually changes from properties of ferrofluid-in-oil emulsion to properties of oil-in-ferrofluid emulsion. Magnetic measurements and microscopic observations made at different values of a magnetic field strength have shown that the magnetic field has no effect on the phase inversion process.

It should be noted that under the action of comparatively strong magnetic field ($\geq 2.5$ kA/m) the coalescence of emulsion droplets was observed. However, in the presented analysis we do not consider the coalescence influence on magnetic properties of an emulsion. Also we have neglected polydispersity of an emulsion.

### 4. Conclusion

Thus, in the present work we studied a soft magnetic system of new type. We have observed that the interaction of ferrofluid emulsions with a magnetic field is accompanied by deformation of emulsion droplets. The phase inversion in the ferrofluid emulsion also has been studied. The influence of these processes on magnetic properties of an emulsion has been investigated. The theory of the observed phenomena has been developed. The emulsions studied can be categorized as fluids with magnetic-field-controllable magnetic properties.

**Acknowledgments.** This work was supported by Russian Foundation for Basic Research (project No. 10-02-90019-Bel_a) and also by Federal Education Agency of the Russian Federation in Scientific Program "Development of Scientific Potential of Higher School".

**Figure Captions**

Fig. 1. Ferrofluid emulsion under the action of magnetic field ($H$=1.5 kA/m). Droplet volume fraction is 10%.

Fig. 2. Semiaxis ratio of oil drops dispersed in ferrofluid as a function of magnetic Bond number, Bm. The volume fraction of oil drops is 10%. Dots are experiments; line is calculations.

Fig. 3. Permeability of the ferrofluid emulsion vs. the magnetic field strength. Dots are experiments; lines are calculations. Squares correspond to the oil-in-ferrofluid emulsion; circles correspond to the ferrofluid-in-oil emulsion. Open symbols correspond to a droplet volume fraction of 20%, filled symbols to 35%.

Fig. 4. Permeability of the ferrofluid vs. the magnetic field strength.

Fig. 5. Permeability of the ferrofluid emulsion vs. the volume fraction of the ferrofluid. Dots are experiments; lines are calculations. Circles correspond to a constant external magnetic field strength $H = 0$; squares to $H = 2.6$ kA/m.

Fig. 6. Calculated dependence of the ferrofluid droplet semiaxis ratio, $\gamma$, on the magnetic Bond number, Bm, for different dispersed phase volume fractions of a ferrofluid-in-oil emulsion: *1*, $\varphi = 0$; *2*, $\varphi = 0.2$; *3*, $\varphi = 0.4$.

Fig. 7. Ratio of the field-induced change in the permeability to the permeability in the absence of field vs. the volume fraction of the ferrofluid. Dots are experiments; lines are calculations. Line *1* correspond to oil-in-ferrofluid emulsion; line *2* to ferrofluid-in-oil emulsion.

Figure 1.

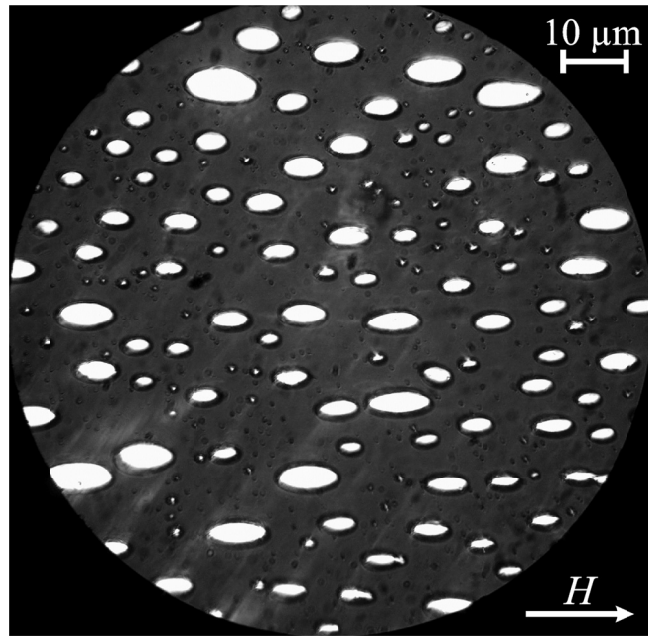

Figure 2.

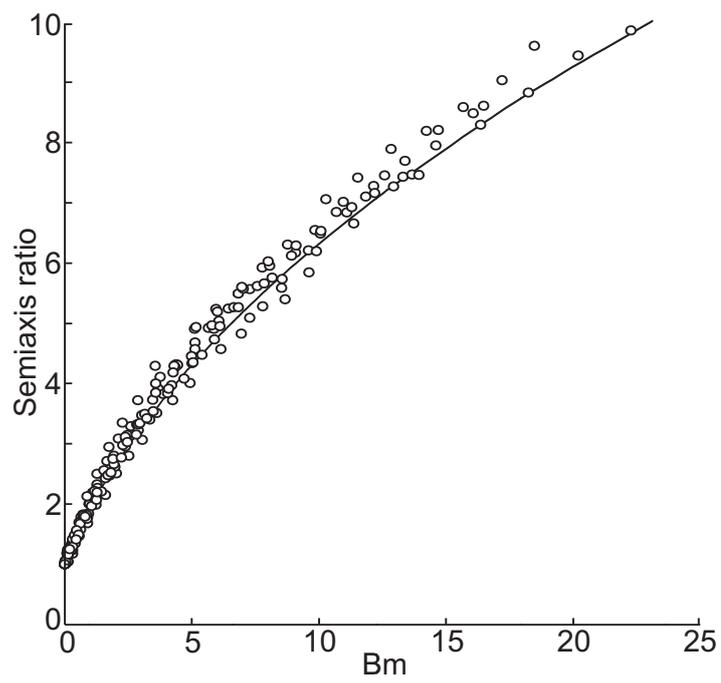

Figure 3.

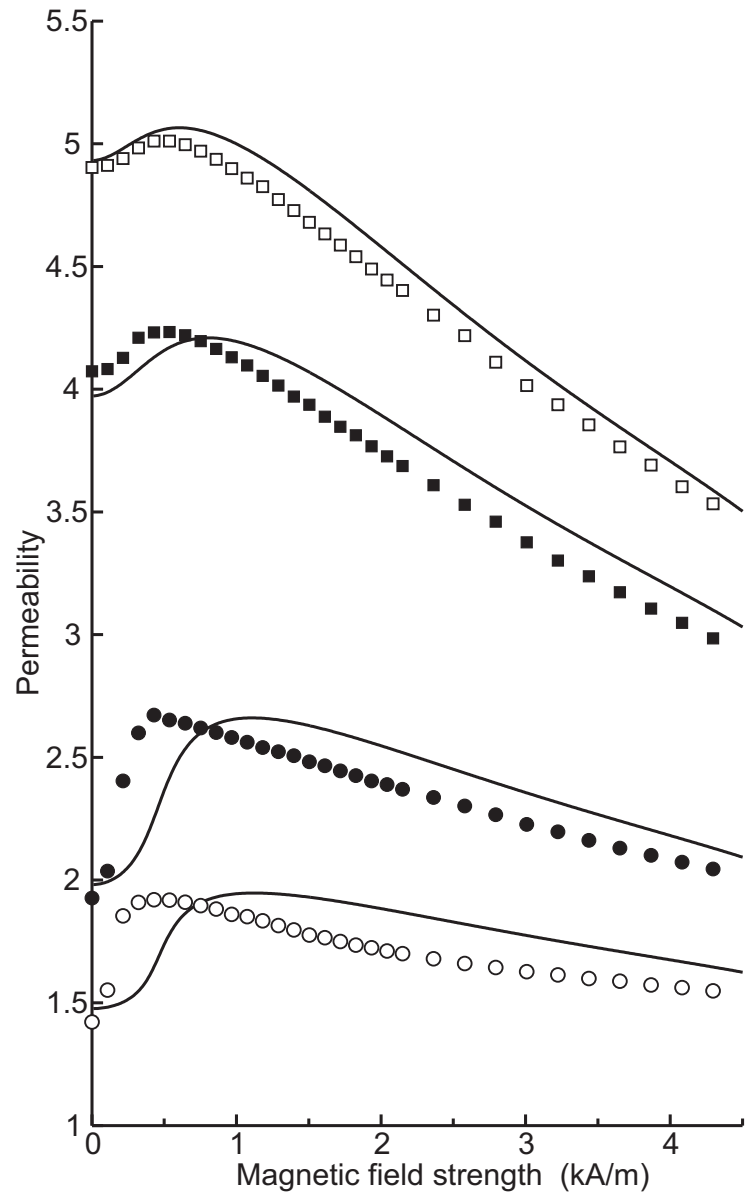

Figure 4.

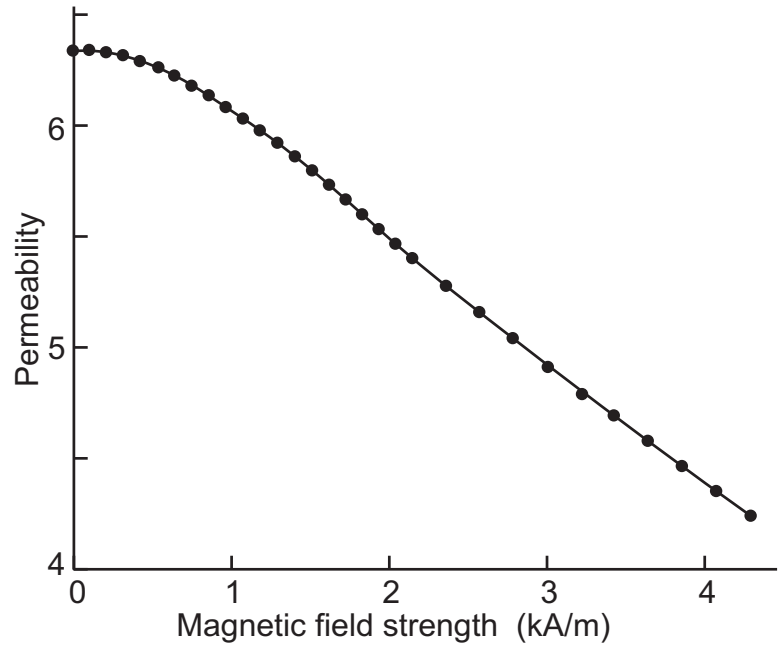

Figure 5.

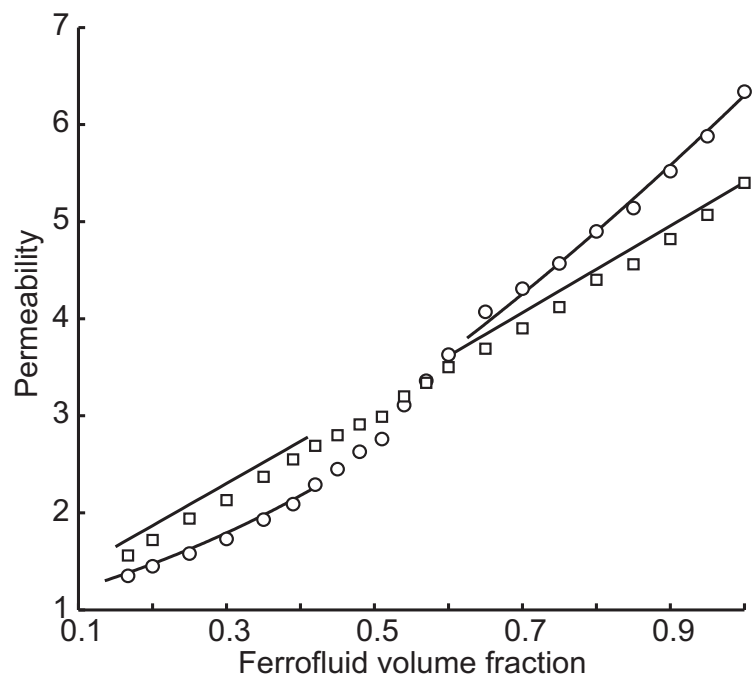

Figure 6.

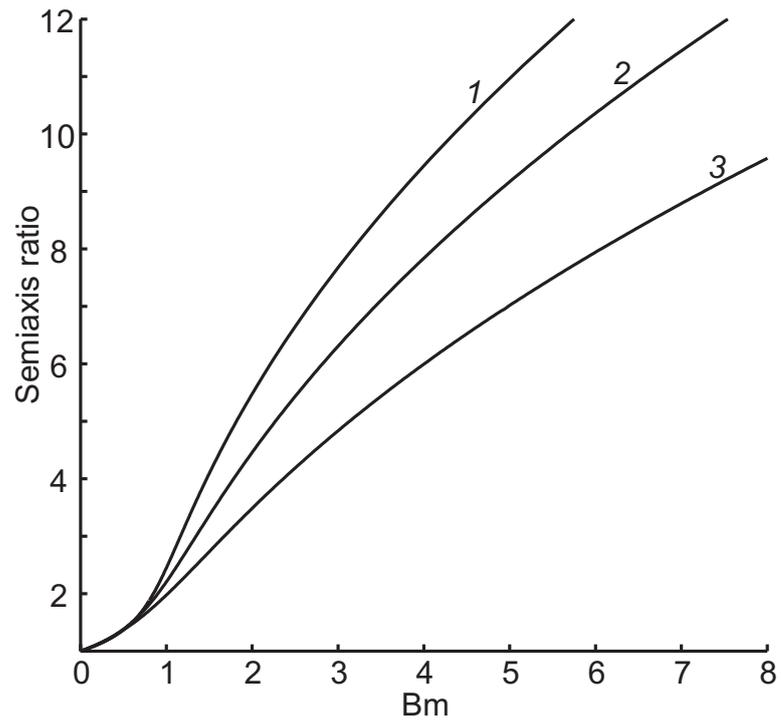

Figure 7.

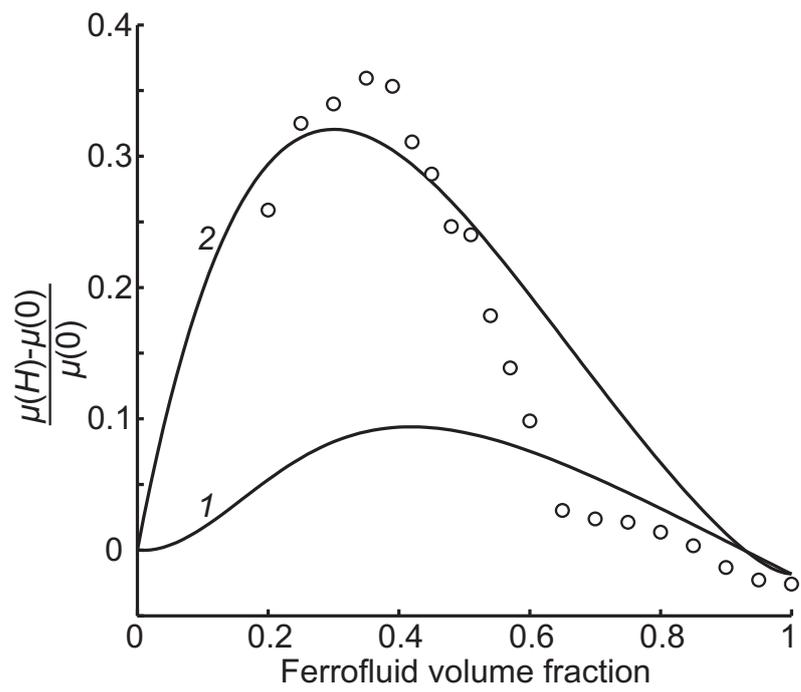